\documentclass[a4paper,amsmath,amssymb,superscriptaddress,aps,prl,reprint,showpacs]{revtex4-1}
\usepackage{color,graphicx,pstricks}
\usepackage{subfigure}
\usepackage{MnSymbol}

\newcommand{\cdag}{c^{\dagger}}
\newcommand{\cnod}{c^{\phantom{\dagger}}}
\newcommand{\e}{\textrm{e}}

\begin{document}
\title{Spinon-orbiton repulsion and attraction mediated by Hund's rule}

\author{Jonas Heverhagen}
\author{Maria Daghofer}
 \affiliation{Institute for Functional Matter and Quantum Technologies, University of Stuttgart,
Pfaffenwaldring 57
D-70550 Stuttgart, Germany}
\affiliation{Center for Integrated Quantum Science and Technology, University of Stuttgart,
Pfaffenwaldring 57
D-70550 Stuttgart, Germany}

\begin{abstract}
  We study the impact of Hund's-rule coupling on orbital excitations,
  as e.g. measured in inelastic resonant x-ray scattering. We find that the
  interpretation in terms of spin-orbit separation, which has been
  derived for one-dimensional systems without Hund's rule, remains robust in its presence.
  Depending on whether or not the orbital excitation includes a spin excitation,
  Hund's rule leads to an attractive or repulsive interaction between
  spinon and orbiton. Attraction and repulsion leave clear signatures through a
  transfer of spectral weight to the lower resp. upper edge of the
  spectrum. 
\end{abstract}
\date{\today} 

\pacs{75.25.Dk,75.30.Ds,75.10.Jm,75.10.Pq} 
%75.25.Dk 	Orbital, charge, and other orders, including coupling of these orders
%75.30.Ds 	Spin waves (for spin-wave resonance, see 76.50.+g)
%75.10.Jm 	Quantized spin models, including quantum spin frustration
%75.10.Pq 	Spin chain models
%71.10.Li 	Excited states and pairing interactions in model systems

\maketitle

\emph{Introduction. }
Low-dimensional quantum systems have long been of special interest due
to the intriguing and often counter-intuitive properties they can
host when quantum fluctuations and interactions come
together. Arguably the strangest concept at play is fractionalization, 
where the electron, an elementary particle, behaves as if it were split
into parts with fractional charges (e.g. in two-dimensional fractional quantum-Hall
states) or into a charge separated from its spin~\cite{Anderson97_fract}. This last concept,
spin-charge separation, applies to an electron or hole 
propagating in a
one-dimensional Mott insulator. Spin and charge  can then be
considered as propagating as 'spinon' and 'holon' with different
velocities, as has been studied theoretically~\cite{PhysRevLett.20.1445,giamarchi} and verified
experimentally using angle-resolved photo-emission spectroscopy~\cite{PhysRevLett.77.4054,BJKim_spincharge_06}. 

More recently, spin-charge separation has been complemented by the
idea of spin-orbit separation~\cite{Wohlfeld:2011hj}. This also occurs in one-dimensional
Mott insulators, but involves an electron being excited into some
unoccupied higher-energy orbital instead of being removed from the
system. As has been pointed out theoretically, the orbital excitation
can then be considered in a manner analogous to a hole and similarly
separates into 'spinon' and 'orbiton'. Experimentally, this has been
verified using resonant inelastic X-ray scattering (RIXS), which can address
orbital excitations, in a cuprate chain compound~\cite{Schlappa:2012hj}. 

The theory behind the analogy of spin-orbit and spin-charge separation rests
on a mapping~\cite{Wohlfeld:2011hj} of the orbital excitation onto a hole-removal excitation that,
strictly speaking, breaks down in the presence of Hund's-rule coupling~\cite{Chen2015}. While
it seems reasonable to assume that small Hund's-rule coupling should not
completely invalidate the scenario, the question of its impact remains
open. As it must be assumed to be present in any realistic material
scenario, we want to assess how far the mapping and the scenario of
spin-orbit separation can be trusted. 

We show in this letter that an interpretation in terms of spinon and orbiton
survives to a very large degree and that the main effect of
Hund's-rule coupling is an interaction between spinon and orbiton. The
issue of spinon-\emph{holon} interaction in the $t$-$J$ model has been 
discussed analytically in the supersymmetric limit~\cite{PhysRevLett.87.177206}, where some exact results can be
obtained, and turned out to be rather subtle~\cite{PhysRevB.71.224424,PhysRevLett.96.059702}. Numerically,
spinon-holon attraction has been followed from the $t$-$J^z$ model,
where it leads to a bound state, to the isotropic $t$-$J$ model, where
it was concluded to be present but too weak for a bound
state~\cite{PhysRevLett.98.266401}. The present work indicates that  
orbital excitations provide an intriguing window into the interactions
between fractionalized excitations: they can address the repulsive
as well as the attractive regime and for strong Hund's-rule coupling,
signatures of spinon-orbiton interaction become quite pronounced.

\emph{Orbital excitations in antiferromagnets.} 
We consider here two orbitals per site, denoted by 1 for the low-energy and 2
for the high-energy state, and the limit
of strong onsite Coulomb repulsion $U$, i.e., we neglect charge
fluctuations.
Second order perturbation theory with intersite hopping $t$ as a small
parameter  $t/U$ then gives a Kugel-Khomskii--type 
model~\cite{KK1982} with the general form 
\begin{align} \label{equ:basics_kk_ham}
   H = 2 \sum^{}_{\langle i ,j\rangle} \left( \vec S_i
   \cdot \vec S_j + \frac{1}{4}  \right) A_{ij} + \sum^{}_{\langle i
     ,j\rangle} K_{ij} + \Delta\sum_{i} T^z_{i}\; ,
\end{align}
where $\vec{S}_i$ describes a spin $S=\tfrac{1}{2}$ at site $i$ and $T_i^z=\tfrac{1}{2}(n_2-n_1)$
is the $z$ component of the orbital pseudospin. Operators $A_{ij}$ and
$K_{ij}$ depend on the orbital degrees of freedom, see below. Bonds
$\langle i ,j\rangle$ run over nearest neighbors (NN), but can be chosen to include longer-range interactions.
Strong crystal field $\Delta\gg t$ ensures that only lower-energy
orbital 1 is occupied  in the ground state.   

Since states with two electrons on one site
enter the perturbation theory as (virtual) intermediate states,
$A_{ij}$ and $K_{ij}$ depend on  the onsite interactions.
It is helpful to first consider any doubly occupied site to have energy $U$,
regardless of its spin and orbital occupation. This neglects processes
like Hund's-rule coupling but brings out the dominant terms  
\begin{align} \label{equ:kk_ham_0}
   A_{ij}^{(0)}  &=\frac{4t^2}{U}\Bigl[
   \bigl( T^z_iT^z_j +\frac{1}{4} \bigr)
   + \frac{1}{2} \left( T_i^+ T_j^- + \textrm{H.c.} \right)\Bigr],\
   K_{ij}^{(0)}  = 0\;.
\end{align}
(We assume identical hopping $t$ in both orbitals.) 
When strong crystal field splitting $\Delta$ 
enforces orbital polarization $T^z_i\equiv -\tfrac{1}{2}$,
the second term in $A_{ij}^{(0)}$ is inactive in the ground state, and
the first term leads to antiferromagnetic (AFM) Heisenberg coupling of the spins
in~(\ref{equ:basics_kk_ham}). 

An orbital excitation is then induced into the AFM state, e.g. by RIXS, which allows for
this excitation to come with or without a spin flip. The two
excitations can be distinguished in
experiment~\cite{1367-2630-13-4-043026,RevModPhys.83.705} and are described by operators 
\begin{align}
T^+(k) &= \frac{1}{\sqrt{L}}\sum_{j,\sigma} \e^{ikj} \cdag_{j,2,\sigma}\cnod_{j,1,\sigma}\quad\textrm{and}\quad\label{eq:orb_exc}\\
(S^x T^+)(k) &= \frac{1}{\sqrt{L}}\sum_{j,\sigma}\e^{ikj} \cdag_{j,2,-\sigma}\cnod_{j,1,\sigma}\;,%= \frac{1}{\sqrt{L}}\sum_{j,\sigma} \sigma\; \e^{ikj}
\label{eq:spinorb_exc}
\end{align}
where $\cdag_{j,2,\sigma}$ ($\cnod_{j,1,\sigma}$) creates
(annihilates) an electron with spin $\sigma=\pm 1=\uparrow,\downarrow$ in the empty 
orbital 2 (occupied orbital 1) on site $j$. $k$ denotes crystal
momentum and runs over the first Brillouin zone of the $L$-site
chain.

The excitations move via the second term in $A_{ij}^{(0)}$, see
Eq.~(\ref{equ:kk_ham_0}), and it turns out that the spin in the upper
orbital 2 is conserved and has no impact on either kinetic or
potential energy~\cite{Kim:2012cr,Wohlfeld:2011hj}. Excitations with
and 
without spin flip are thus equivalent and
can be mapped onto a spinless hole moving
in an AFM background. This mapping between orbital excitations and hole
dynamics has been used extensively to analyze RIXS in
one-dimensional cuprates~\cite{Schlappa:2012hj,PhysRevB.88.195138,PhysRevLett.114.096402} and two-dimensional iridates~\cite{Kim:2012cr,Kim:2014ku,PhysRevLett.116.106401}.

\begin{figure*}[bt]
  \subfigure{\includegraphics[width=\columnwidth,trim=0 115 0 55, clip]
    {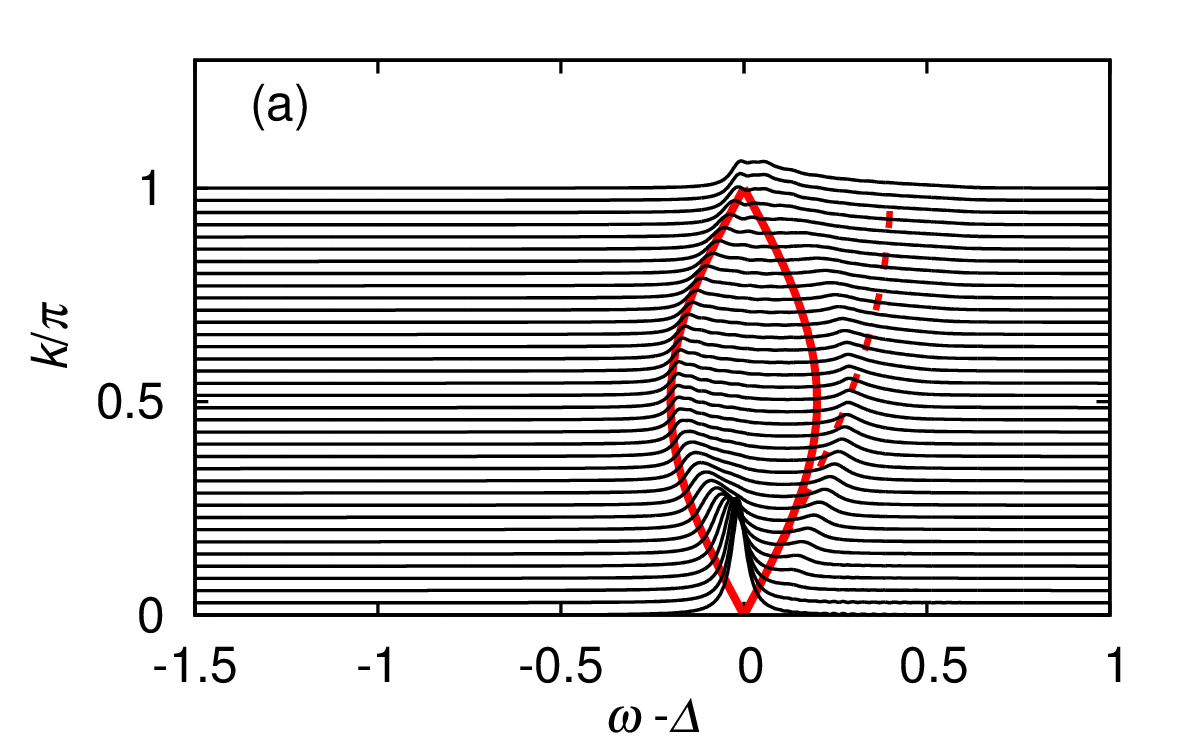}\label{fig:spectra_T_1}}\hfill
  \subfigure{\includegraphics[width=\columnwidth,trim=0 115 0 55, clip]
    {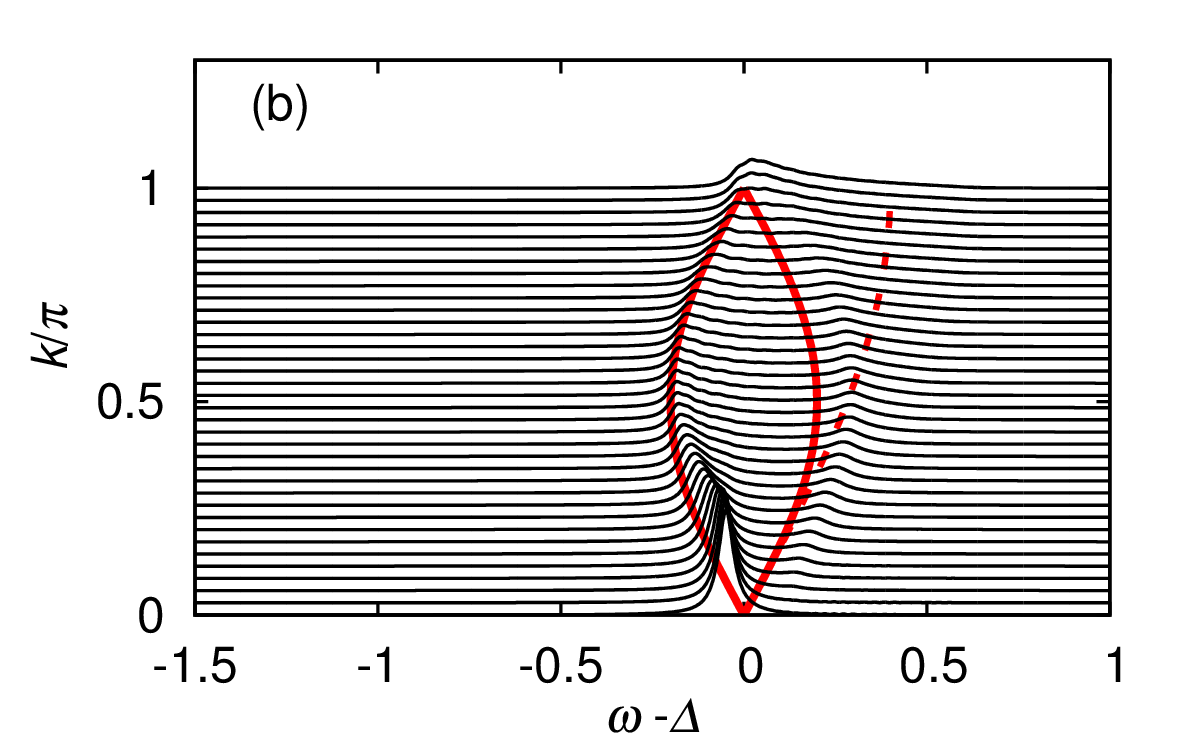}\label{fig:spectra_ST_1}}\\[-2em]
  \subfigure{\includegraphics[width=\columnwidth,trim=0 115 0 0, clip]
    {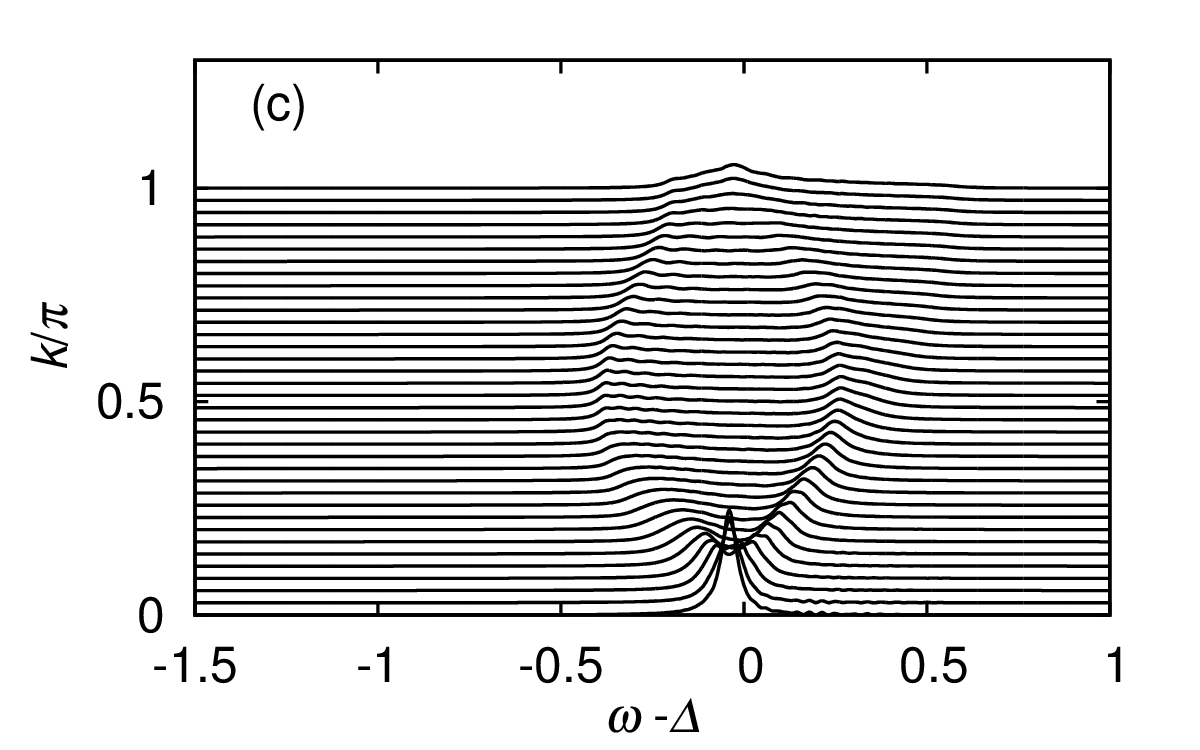}\label{fig:spectra_T_3}}\hfill
    \subfigure{\includegraphics[width=\columnwidth,trim=0 115 0 0, clip]
    {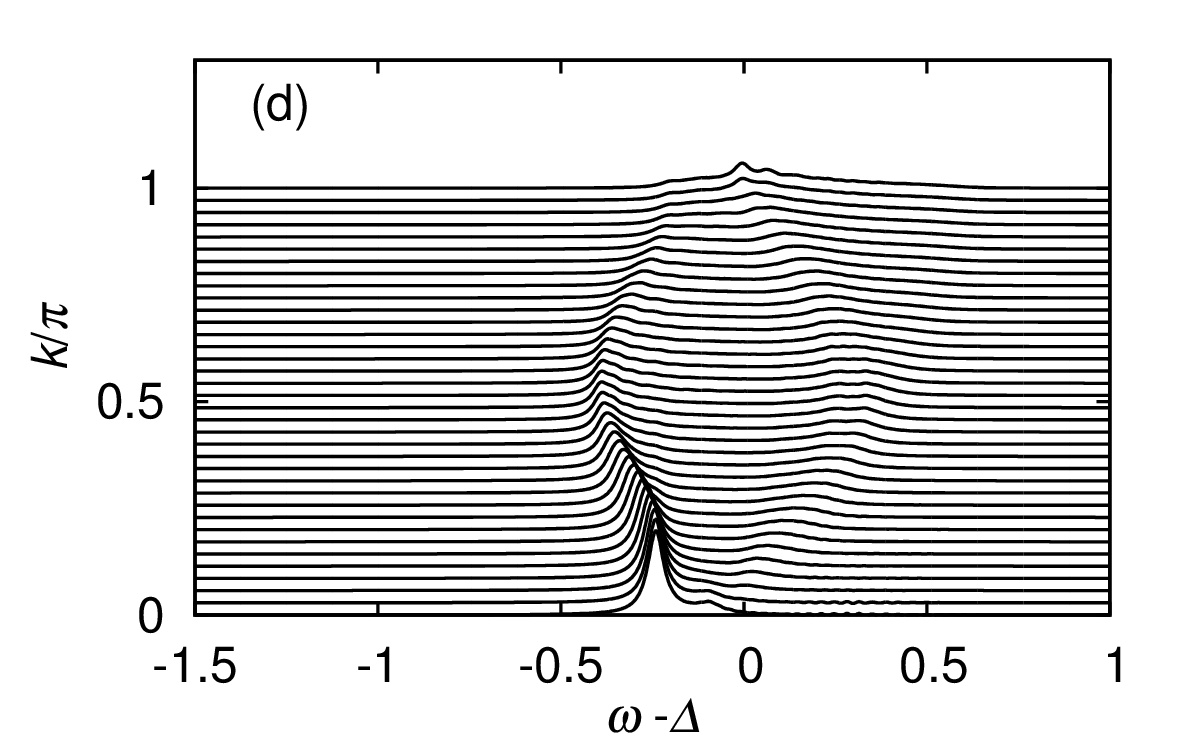}\label{fig:spectra_ST_3}}\\[-2em]
  \subfigure{\includegraphics[width=\columnwidth,trim=0 0 0 0, clip]
    {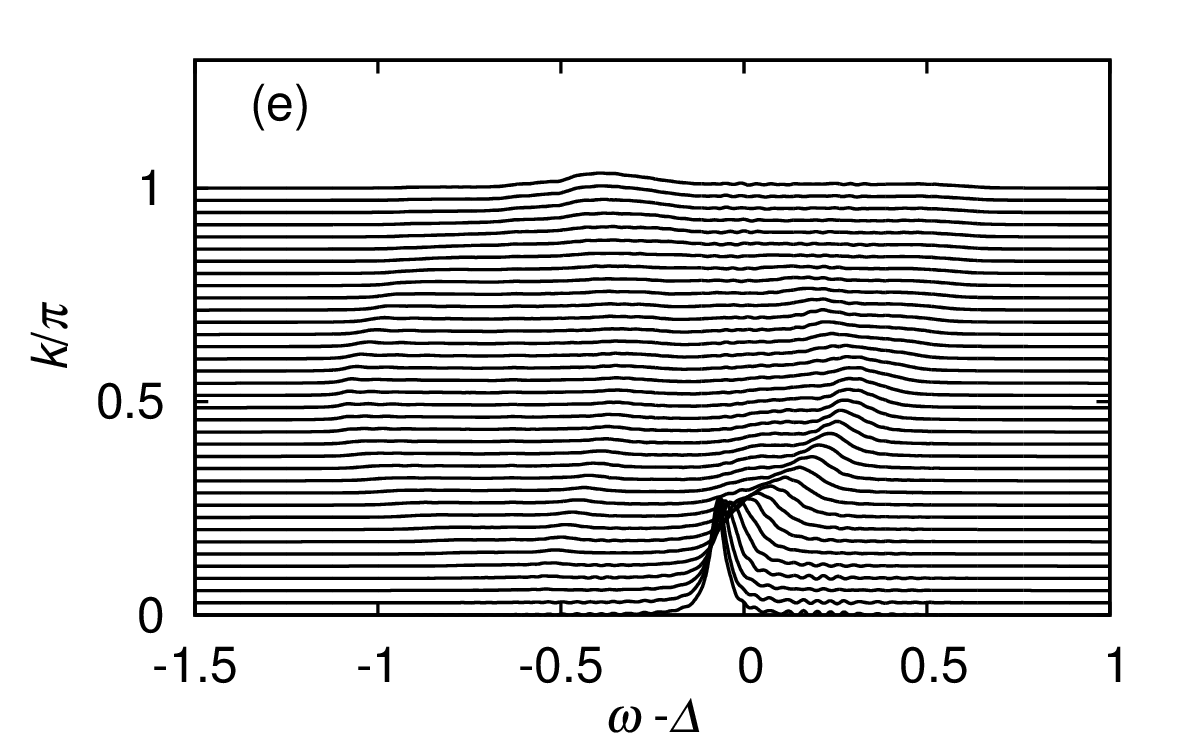}\label{fig:spectra_T_5}}\hfill
  \subfigure{\includegraphics[width=\columnwidth,trim=0 0 0 0, clip]
    {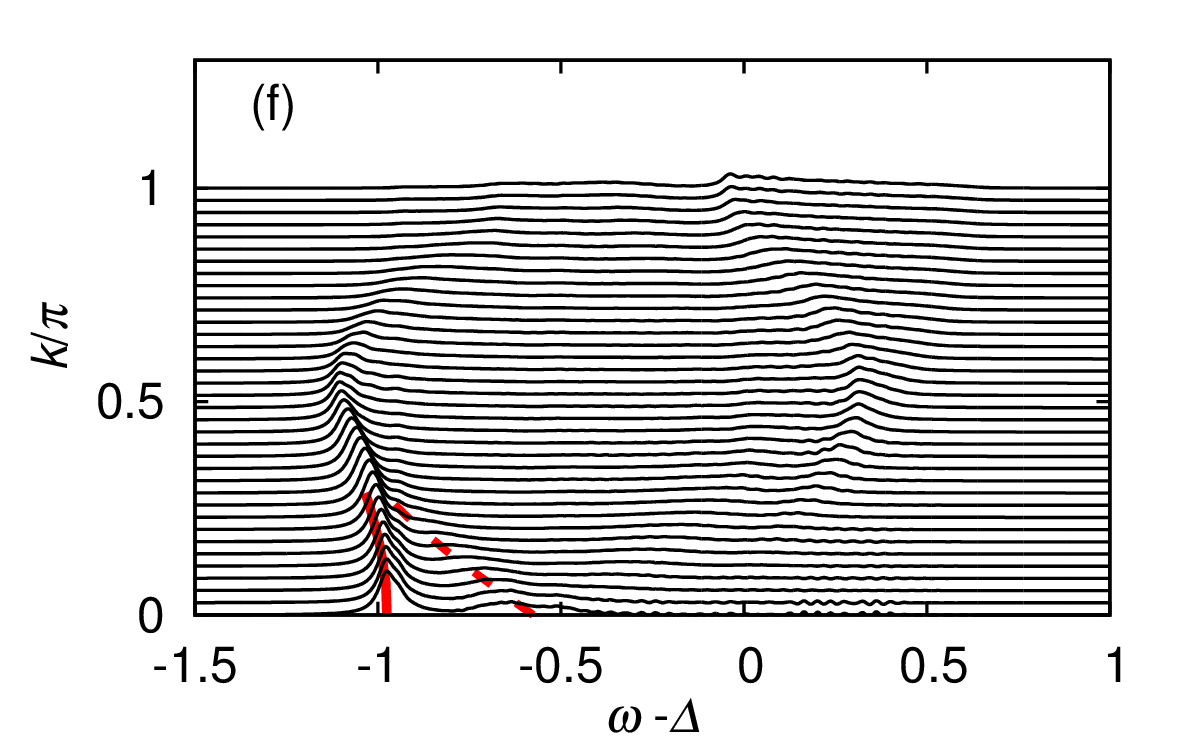}\label{fig:spectra_ST_5}}\\[-1em]
\caption{Orbital excitations with increasing $J_H=J_P=\Delta
  U/2$. Spectra for the pure orbital excitation Eq.~(\ref{eq:orb_exc}) without a
  spin flip are shown in the left column for \subref{fig:spectra_T_1} $J_H=t=U/20$, \subref{fig:spectra_T_3}
  $J_H=3t=3U/20$, and \subref{fig:spectra_T_5} $J_H=5t=U/4$. The right column
  gives spectra for the spin-orbital excitation
  Eq.~(\ref{eq:spinorb_exc}), which includes a spin flip, for \subref{fig:spectra_ST_1} $J_H=t$, \subref{fig:spectra_ST_3}
  $J_H=3t$, and \subref{fig:spectra_ST_5} $J_H=5t$. The broad solid (dashed)
  lines in \subref{fig:spectra_T_1} and  \subref{fig:spectra_ST_1}
  give the approximate support of the one-spinon--one-holon 
  (three-spinon--one-holon) part of the one-particle spectrum of the
  $t$-$J$ model~\cite{PhysRevLett.73.2887,PhysRevLett.81.5402}. This
  support is consistent with a phenomenological analysis in terms of
  spinon and holon interacting via a phase
  string~\cite{PhysRevB.56.3548} and corresponds to the orbital
  excitation spectrum for $J_H=0$. In
  \subref{fig:spectra_ST_5}, the broad solid and dashed lines are guides to
  the eye following the peaks of the branches identified as 'spinon'
  and 'orbiton' branches.  
  Results obtained
  with spin CPT based on $L=24$ sites, Coulomb repulsion $U=20 t$,
  crystal field $\Delta=10 t$. \label{fig:spectra_T_TS}} 
\end{figure*}

\emph{Richer structure of the doubly occupied site.} 
While the dominant terms discussed above are helpful and illustrative,
a realistic description of two electrons on one site has to take into
account processes beyond pure charge interactions. Let $U$ denote
Coulomb repulsion felt by two electrons in the 
same orbital on the same site $i$. Due to reduced
overlap of the wave functions, their interaction  $U' = U-\Delta U <
U$ is weaker if they occupy different orbitals. In that case, Hund's-rule coupling $-2J_H\vec{S}_{i,1}\vec{S}_{i,2}$ moreover favors their ferromagnetic (FM) alignment. Finally, a 'pair
hopping' $J_P$  involves a doubly occupied high-energy
orbital 2 and is suppressed here by the large crystal field. The full Kugel-Khomskii Hamiltonian is given by
\begin{align} 
  & A_{ij}  = \frac{4Ut^2}{U^2-J_P^2} \left( T^z_iT^z_j +\frac{1}{4}
   \right)
   +\frac{4J_Ht^2}{U'^2-J_H^2} \left( T^z_iT^z_j -\frac{1}{4} \right)
   + \nonumber \\
& + \frac{2U't^2 }{U'^2-J_H^2}  \left( T_i^+ T_j^- + \textrm{H.c.} \right)
 -\frac{2J_P t^2}{U^2-J_P^2} \left( T_i^-T_j^- + \textrm{H.c.} \right)\label{equ:kk_ham_A}
\end{align}
and
\begin{align} 
  & K_{ij}  = -\frac{4Ut^2 }{U^2-J_P^2} \left( T^z_iT^z_j +\frac{1}{4}
   \right)   +\frac{4U't^2}{U'^2-J_H^2} \left( T^z_iT^z_j -\frac{1}{4}
   \right) + \nonumber \\
&+ \frac{2J_Ht^2 }{U'^2-J_H^2} \left( T_i^+ T_j^- + \textrm{H.c.} \right)
+\frac{2J_P t^2}{U^2-J_P^2} \left( T_i^-T_j^- + \textrm{H.c.} \right). \label{equ:kk_ham_K}
\end{align}
We use here relations $J_H=J_P$ and $\Delta U=2J_H$, which arise
naturally for symmetry-related
orbitals~\cite{griffith,Castellani:1978p1292}, but have checked that
deviations do not significantly alter our results. 

We apply (Lanczos) exact diagonalization to Hamiltonian~(\ref{equ:basics_kk_ham}) with orbital
operators~(\ref{equ:kk_ham_A})-(\ref{equ:kk_ham_K}). To reach longer chains, 
only states with at most one electron in the higher-energy
orbital 2 are kept, which does not affect results in our limit of large crystal-field
splitting. Exact diagonalization is complemented with
spin--cluster-perturbation theory~\cite{Ovchinnikov2010}, which gives
limited access to momentum points not directly available on the
directly solved cluster and which has been previously applied to orbital
excitations~\cite{Chen2015}.

\emph{Numerical results and spinon-orbiton interaction.} 
Figure~\ref{fig:spectra_T_TS} shows spectra for
excitations (\ref{eq:orb_exc}) and (\ref{eq:spinorb_exc}) without and
with a spin flip, for increasing deviation from the high-symmetry case
(\ref{equ:kk_ham_0}). At small $J_H/U=1/20$, excitations with and without
spin flip look nearly identical, see Figs.~\ref{fig:spectra_T_1}
and~\ref{fig:spectra_ST_1}. They also strongly resemble the $J_H=0$
result~\cite{Wohlfeld:2011hj}, which in turn correspond to the
one-particle spectral density of the supersymmetric $t$-$J$ model~\cite{PhysRevB.87.195105} with
$t=J/2$. The spectrum can then be described in terms of spinon and
holon interacting via a phase string~\cite{PhysRevB.56.3548}, where
the role of the holon is here taken by the 'orbiton'. The lens-shaped dominant feature can thus be identified with
the one-spinon--one-holon (orbiton) part of the spectrum, while the
additional small weight at higher energy towards $k=\pi$ comes from
states with three
spinons~\cite{PhysRevLett.73.2887,PhysRevLett.81.5402}.  

At larger $J_H/U=3/20$, the lens can still be recognized, albeit with a
broadened energy range. However, spectral weight has clearly shifted
to its high-energy (low-energy) edge for the pure orbital (combined
spin-orbital) excitation, see Figs.~\ref{fig:spectra_T_3}
and~\ref{fig:spectra_ST_3}. Finally at  $J_H/U=5/20=1/4$, energy range
has further increased and spectral weight is almost completely
located on the upper (lower) side without (with) a spin flip. For the
combined spin-orbital excitation, Fig.~\ref{fig:spectra_ST_5} shows
features like the 'spinon' and 'holon' branches familiar from the
$t$-$J$ model, however, the 'holon' is broadened.

In order to interpret the features and understand their origin,
corrections to  (\ref{equ:kk_ham_0}) in first-order of $\tfrac{1}{U}$
can be analyzed. The part $K_{ij}$ decoupled from spins  no longer vanishes, 
\begin{align} 
  K_{ij}^{(1)} & = \frac{4t^2}{U}\Bigl[
    \frac{\Delta U}{U}\bigl( T^z_iT^z_j -\frac{1}{4} \bigr) +   \frac{J_H}{2U}\bigl( T_i^+ T_j^- + \textrm{H.c.} \bigr)\nonumber\\
    &\quad 
    +\frac{J_P}{2U} \bigl( T_i^-T_j^- + \textrm{H.c.}\bigr) \Bigr]\;,\label{equ:kk_ham_KJ}
\end{align}
where the last term $\propto J_P$ is suppressed by crystal-field
splitting $\Delta$ and the first
term $\propto \Delta U$ gives a small overall energy shift. The second term $\propto J_H$ allows the excited orbital to
move without an \emph{onsite} spin flip, so that the spin of the \emph{excitation} can 
flip. In the 1D chain, this can induce additional spinons and is a likely
reason for, e.g., the broadening of the 'holon' 
branch in  Fig.~\ref{fig:spectra_ST_5}. Despite its rather minor role here,
we expect this term to have a more decisive effect in higher 
dimensions, where it would allow the orbital excitation to travel
'freely' through an AFM ordered state without creating a string potential. 

Corrections to $A_{ij}$ are 
\begin{align} 
   A_{ij}^{(1)}  &=\frac{4t^2}{U}\Bigl[
\frac{J_H}{U}\bigl( T^z_iT^z_j -\frac{1}{4} \bigr)+\frac{\Delta U}{2U}\left( T_i^+ T_j^- + \textrm{H.c.} \right) \nonumber \\
   &\quad 
-\frac{J_P}{2U}\left( T_i^-T_j^- + \textrm{H.c.}
\right)\Bigr]\;,\label{equ:kk_ham_AJ}
\end{align}
where the last term is again suppressed. The second terms here and in
Eq.~(\ref{equ:kk_ham_0}) have exactly the same form and same sign, so that the
main effect of $U'<U$ is to increase orbiton hopping relative to spin
superexchange~\cite{Kim:2012cr}. This in turn increases band width and makes
the orbiton faster than the spinon, so that we recover the 'usual'
spinon-holon scenario in Fig.~\ref{fig:spectra_ST_5}. 

\begin{figure}
  \subfigure{\includegraphics[width=0.48\columnwidth,trim=200 160 225 125, clip]{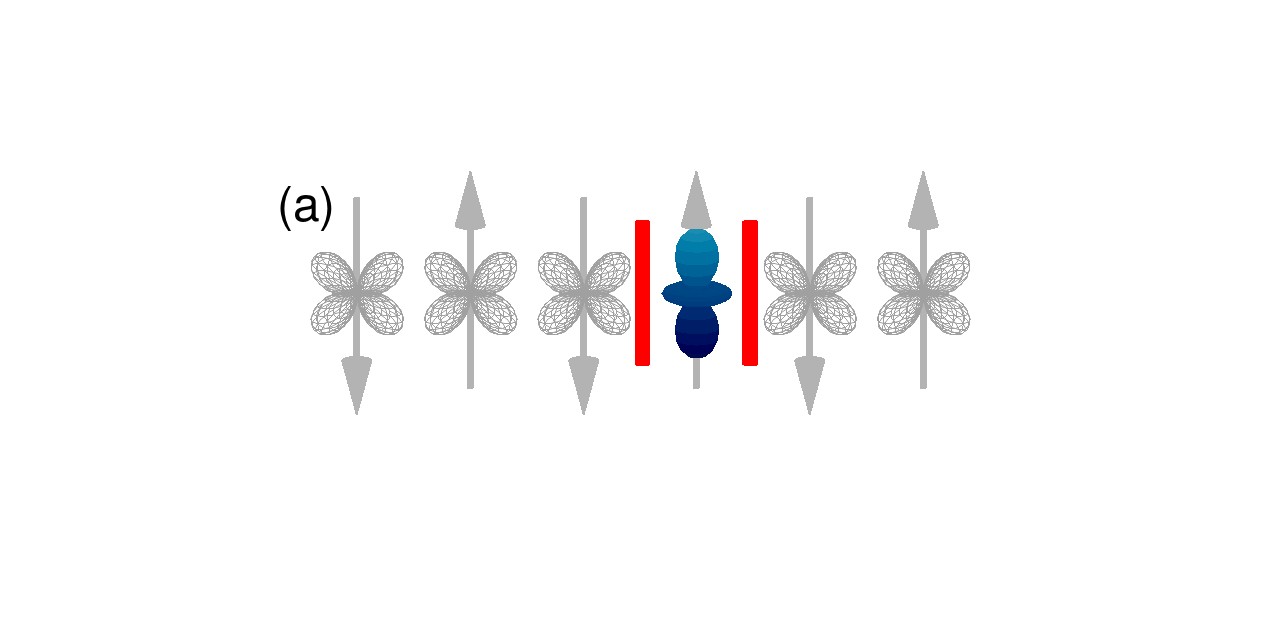}\label{fig:cartoon_Tx0}}\hfill
  \subfigure{\includegraphics[width=0.48\columnwidth,trim=200 160 225 125, clip]{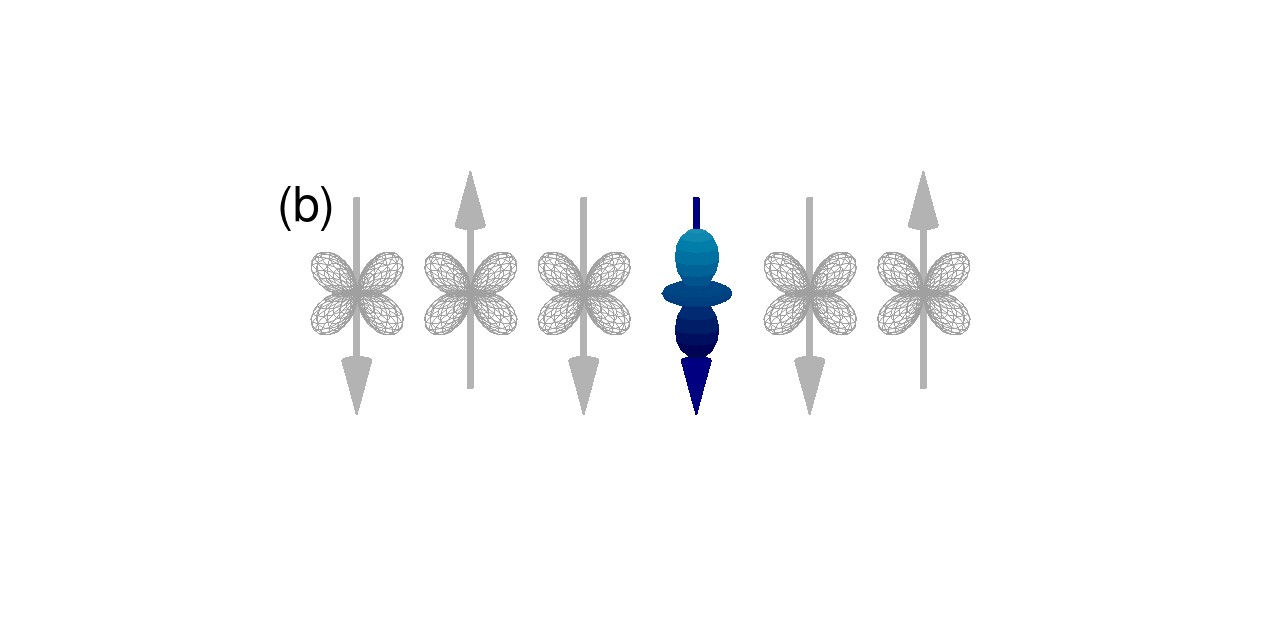}\label{fig:cartoon_SxTx0}}\\
  \subfigure{\includegraphics[width=0.48\columnwidth,trim=200 160 225 125, clip]{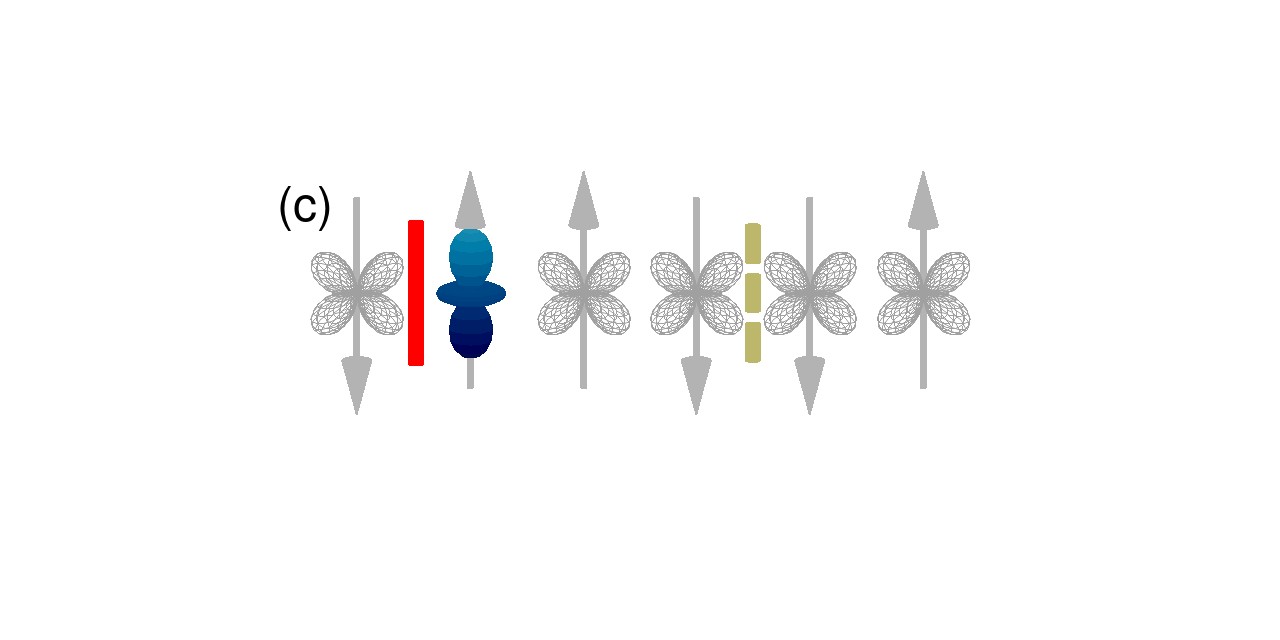}\label{fig:cartoon_Txfar}}\hfill
  \subfigure{\includegraphics[width=0.48\columnwidth,trim=200 160 225 125, clip]{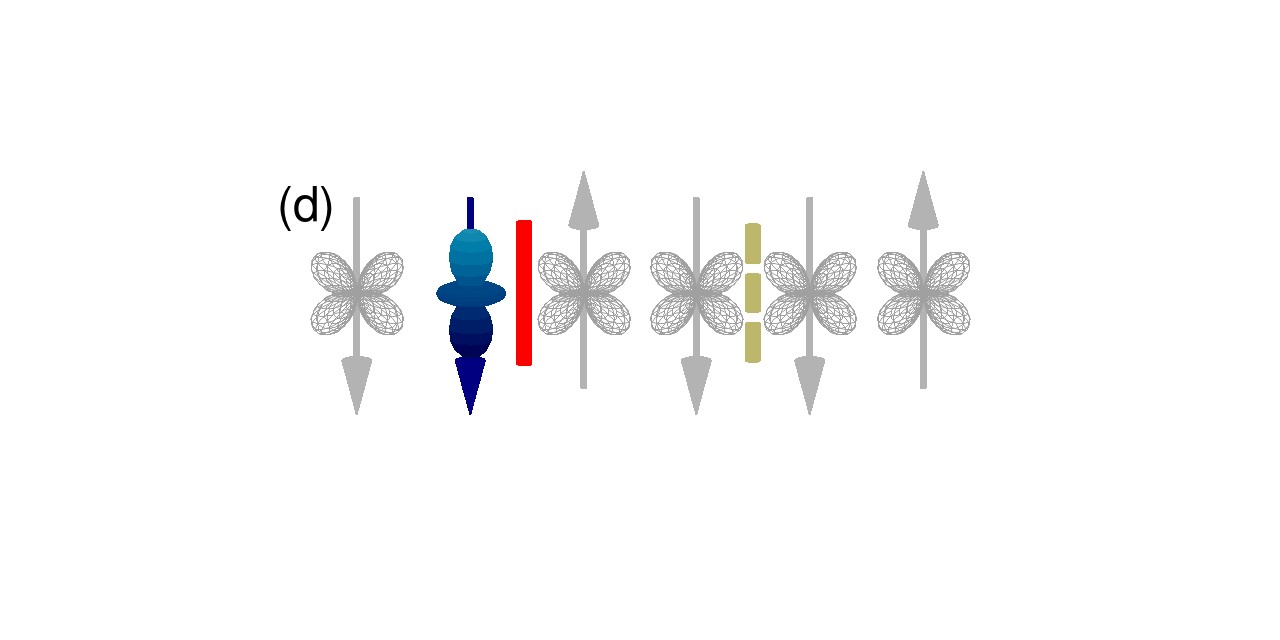}\label{fig:cartoon_SxTxfar}}
\caption{Cartoon of the effective spinon-orbiton
  repulsion/attraction. In \subref{fig:cartoon_Tx0} and
  \subref{fig:cartoon_SxTx0}, orbital excitations without/with spin
  flip are created, spinon and orbiton are located on the same
  spot.
  In \subref{fig:cartoon_Txfar} and
  \subref{fig:cartoon_SxTxfar}, the orbiton has moved two sites, leaving
  behind the spinon.
  Broken vertical bars between sites indicate spinons; solid bars
  violations of Goodenough-Kanamori rules on bonds with alternating
  orbitals.
  \label{fig:cartoon}} 
\end{figure}

Finally, the first term of (\ref{equ:kk_ham_AJ}) becomes negative
between two sites with different orbital occupation because
$T_i^zT_j^z=-\tfrac{1}{4}$ in that case, while the term vanishes
for  identical orbitals and $T_i^zT_j^z=+\tfrac{1}{4}$. Negative
$A_{ij}$ implies FM spin-spin coupling in the Kugel-Khomskii
Hamiltonian (\ref{equ:basics_kk_ham}). This is opposite to the AFM
coupling between identical orbitals that comes from the first term
of~(\ref{equ:kk_ham_0}), which in turn vanishes between sites with
different orbital occupation. The sign change of the magnetic
interaction driven by different orbital occupation, known as
Goodenough-Kanamori rules~\cite{PhysRev.100.564,1959JPCS1087Kanamori},
often contributes 
to complex magnetic orderings in the presence of orbital degrees of freedom. We argue here that this effect also 
mediates an interaction between the excited orbital and the spinon,
see Fig.~\ref{fig:cartoon}, that plays out differently for the
excitations with and without a spin flip.

In (short-range) AFM order, the excited electron after a
process Eq.~(\ref{eq:spinorb_exc}) with spin flip has the same spin as its
two neighbors in lower-energy orbitals, see Fig.~\ref{fig:cartoon_SxTx0}, exactly the situation
energetically favored by the first term of (\ref{equ:kk_ham_AJ}). If the excited orbital (i.e. the orbiton) moves 
away, leaving behind a domain wall (i.e. the spinon), it is generally
found between spins of opposite sign, so that only one bond can profit
from FM exchange. The other bond,
indicated by a red vertical bar in Fig.~\ref{fig:cartoon_SxTxfar}, 
is AFM and raises the energy for separating spinon and orbiton; orbiton and spinon thus see an attractive
potential. In contrast, an orbital excited \emph{without} a spin flip
in process Eq.~(\ref{eq:orb_exc}) has spin opposite to its two
neighbors, see Fig.~\ref{fig:cartoon_Tx0}, and both bonds pay
energy. Separated from the spinon and sitting between an
up and a down spin, it has spin parallel to one of its neighbors, see
Fig.~\ref{fig:cartoon_SxTxfar}, which reduces energy cost. Accordingly, spinon and orbiton repel
each other here.

However, the cartoon Fig.~\ref{fig:cartoon} with its perfectly ordered
Ising spins overemphasizes spinon-orbiton attraction, because it
suggests that the domain wall costs energy
$\propto J=\tfrac{4t^2}{U}$
anywhere except at the site of the orbiton. This interaction,
indicated by broken vertical bars in Fig.~\ref{fig:cartoon}, 
would be independent of $J_H$ and  
indeed binds spinon and holon together in the $t$-$J^z$ model~\cite{PhysRevB.57.6444}. 
While the effect has been found too weak to induce a
bound state in the spin-isotropic $t$-$J$ model with its half-filled
spinon-sea groundstate, it is sizable on finite
chains~\cite{PhysRevLett.98.266401}.

Fortunately, Hund's-rule--driven spinon-orbiton interaction can be distinguished
from this 'baseline' interaction even on small systems by use of open boundary conditions (OBC). To do so, we set the 
crystal field splitting to a negative value at one site $i_2=\tfrac{L}{2}$ near the
center of an OBC chain. In the ground state, orbital 2 thus has one
electron at site $i_2$ and the AFM state in orbital 1 has at least one domain wall, which can
sit either around site $i_2$ or at the open ends. At $J_H=0$, both
positions have equal weight. In the presence of $J_H>0$, the preferred position depends on total
$S^z=\tfrac{1}{2}(N_\uparrow-N_\uparrow)$: For $S^z=1$ (describing the
case with a spin flip) the domain wall is found predominantly around $i_2$ while it
prefers the open chain ends for $S^z=0$, indicating attraction
resp. repulsion. Numerical spectra in Fig.~\ref{fig:spectra_T_TS}
do not show bound states, but the build-up of spectral weight at 
lower (higher) excitation energies can be explained by including such a
spinon-holon interaction into a phenomenological
description~\cite{PhysRevB.56.3548} in terms of spinon and holon.

\emph{Conclusions.}

Fractionalization of the electron into spin and charge has long been
realized as an intriguing property of one-dimensional
systems. The question of interactions between the fractionalized parts
then naturally arises, even if they are not be strong enough to glue the
electron back together. We find here that orbital excitations can
offer insights into this aspect of  spin-charge separation that are
not easily accessible to one-particle excitations.

Orbital excitations had been shown to exhibit spin-orbit separation in
analogy to spin-charge separation, with the orbiton taking the role of
the holon. We have here seen that Hund's rule leads to an attraction
or repulsion between spinon and orbiton, depending on whether the
excitation includes a spin flip or not. Their microscopic origin can
be understood as a dynamic signature of the Goodenough-Kanamori rules
that favor FM (AFM) spins on bonds with different (identical)
orbitals.  Hund's-rule--induced
interactions are not strong enough to
induce (anti-)bound states, but they lead to clear signatures by shifting spectral weight
to the upper resp. lower edge of the one-spinon--one-orbiton part of
the spectrum.

\begin{acknowledgments}
We thank K. Wohlfeld for stimulating discussions and careful reading of the manuscript.
This research was supported by the Deutsche Forschungsgemeinschaft,
via the Emmy-Noether program (DA 1235/1-1) and FOR1807 (DA 1235/5-1).
\end{acknowledgments}

%\bibliography{orbital,methods,general} 

\end{document}